\documentclass[pss,fleqn]{w-art}
\usepackage{times}
\usepackage{w-thm}
\usepackage[]{graphicx}
\usepackage{cite}
\begin{document}
\DOIsuffix{theDOIsuffix}
\Volume{XX} \Issue{1} \Month{01} \Year{2004}
\pagespan{1}{}
\Receiveddate{1 Jan 2004} \Reviseddate{1 Jan 2004} \Accepteddate{1 Jan 2004} \Dateposted{1 Jan
2004}
\keywords{phase transition, critical point, disorder} \subjclass[pacs]{05.70.Jk, 64.60.Ak, 75.40.-s}



\title[Critical points and quenched disorder]{Critical points and quenched disorder: From Harris criterion to
rare regions and smearing}


\author[]{Thomas Vojta\footnote{Corresponding
     author: e-mail: {\sf vojtat@umr.edu}, Phone: +1\,573\,341\,4793,
     Fax: +1\,573\,341\,4715}}
\author[]{Rastko Sknepnek}
\address[]{Department of Physics, University of Missouri-Rolla, MO 65401, USA}
\dedicatory{Dedicated to Prof. Michael Schreiber on the occasion of his 50th birthday}
\begin{abstract}
We consider the influence of quenched spatial disorder on phase transitions in classical and
quantum systems. We show that rare strong disorder fluctuations can have dramatic effects on
critical points. In classical systems with sufficiently correlated disorder or in quantum
systems with overdamped dynamics they can completely destroy the sharp phase transition by
smearing. This is caused by effects similar to but stronger than Griffiths phenomena: True
static order can develop on a rare region while the bulk system is still in the disordered
phase. We discuss the thermodynamic behavior in the vicinity of such a smeared transition using
optimal fluctuation theory, and we present numerical results for a two-dimensional model
system.

\end{abstract}
\maketitle                   

\section{Introduction}
The effects of impurities, dislocations, grain boundaries or other types of quenched disorder
on phase transitions and critical points have fascinated physicists for more than 3 decades. In
the early days, it was thought that quenched disorder destroys any critical point, because in
the presence of defects, the system divides itself up into spatial regions which independently
undergo the phase transition at different temperatures. At such a smeared phase transition, a
unique critical temperature for the entire system does not exist, and the singularities in the
thermodynamic quantities, which are the characteristic feature of a critical point, are rounded
(see Ref.\ \cite{Grinstein} and references therein).
However, subsequently it became clear that generically a phase transition remains sharp in the
presence of quenched disorder, at least for classical systems with short-range disorder
correlations. The fate of a particular clean critical point under the influence of impurities is
controlled by the Harris criterion \cite{Harris74}: If the correlation length critical exponent
$\nu$ fulfills the inequality $\nu\ge2/d$ where $d$ is the spatial dimensionality, the disorder
does not affect the critical behavior. If the Harris criterion is violated, the generic result
in classical systems is a new critical point with conventional power law scaling but new
exponents which fulfill the Harris criterion.

In recent years, the problem of quenched disorder and critical points has reattracted a lot of
attention because new results have challenged the simple classification outlined above. At
quantum phase transitions, i.e., transitions occurring at zero temperature as a function of a
nonthermal control parameter, quenched disorder can lead to exotic critical points where the
conventional power-law scaling is replaced by exponential (activated) scaling (see Ref.
\cite{Fisher9295} for a prototypical example, the random transverse field Ising model). Similar
effects can also be found in classical systems with linear defects \cite{McCoyWu}. Recently,
even more dramatic effects have been proposed to occur at some quantum phase transitions in
itinerant electronic systems \cite{rounding_prl}. At these transitions a true static order
parameter can develop on spatial regions that are more strongly coupled than the bulk system
because of disorder fluctuations. Thus, these rare regions undergo a phase transition
independently from the rest of the system, leading to a smeared global transition exactly in
the sense discussed at the beginning of this section.

In this paper we explore this smeared phase transition scenario in more detail. We show that
disorder-induced smearing of a critical point is a ubiquitous phenomenon occurring in a wide
variety of systems ranging from the above-mentioned itinerant electronic systems to classical
Ising magnets with planar defects and to non-equilibrium phase transitions. To illustrate the
smearing mechanism, we present computer simulations of a two-dimensional (2D) model system. The
paper is organized as follows: In Section \ref{sec:cps}, we discuss scaling scenarios for a
critical point in the presence of quenched disorder. Section \ref{sec:griffiths} is devoted to
rare regions and Griffiths singularities. In Section \ref{sec:smearing} we discuss how these
rare regions can destroy a critical point, and we present numerical results in Section
\ref{sec:numerics}.


\section{Critical point scenarios}
\label{sec:cps}

In this section, we discuss possible scaling scenarios for (sharp) critical points in the
presence of weak quenched disorder.  Weak disorder refers to types of defects which
introduce spatial variations of the coupling strength but no frustration or random external
fields. For these systems, three broad classes can be distinguished according to the behavior
of the disorder under coarse graining.

The first class consists of clean critical points that fulfill the Harris criterion
\cite{Harris74} $\nu\ge 2/d$. At these transitions, the disorder {\em decreases} under coarse
graining, and the system becomes asymptotically homogeneous at large length scales.
Consequently, the critical behavior of the dirty system is identical to that of the clean
system. Technically, this means the disorder is renormalization group irrelevant, and the clean
renormalization group fixed point is stable. In this first class the macroscopic observables
are self-averaging at the critical point, i.e., the relative width of their probability
distributions vanishes in the thermodynamic limit \cite{AharonyHarris96,WisemanDomany98}. A
prototypical example in this class is the 3D classical Heisenberg model whose clean correlation
length exponent is $\nu\approx 0.698$ (see, e.g., \cite{Janke93}), fulfilling the Harris
criterion.

The other two classes occur if the clean critical point violates the Harris criterion. In the second class, the
system remains inhomogeneous at all length scales with the relative strength of the
inhomogeneities approaching a finite value for large length scales. The resulting critical
point still displays conventional power-law scaling but with new critical exponents which
differ from those of the clean system (and fulfill the Harris criterion). These transitions are
controlled by renormalization group fixed points with finite disorder. Macroscopic observables
are not self-averaging, the relative width of their probability distributions approaches a
size-independent constant \cite{AharonyHarris96,WisemanDomany98}. An example in this second
class is the classical 3D Ising model. Its clean correlation length exponent, $\nu\approx
0.627$ (see, e.g. \cite{Ferrenberg91}) does not fulfill the Harris criterion. Introduction of
quenched disorder, e.g., via dilution, thus leads to a new critical point with an exponent of
$\nu\approx 0.684$ \cite{Ballesteros98}.

At critical points in the third class, the relative magnitude of the inhomogeneities {\em
increases} without limit under coarse graining. The corresponding renormalization group fixed
points are characterized by infinite disorder strength. At these infinite-randomness critical
points, the power-law scaling is replaced by activated (exponential) scaling. The probability
distributions of macroscopic variables become very broad (even on a logarithmic scale) with the
width diverging with system size. Consequently, averages are often dominated by rare events,
e.g., spatial regions with atypical disorder configurations. This type of behavior was first
found in the McCoy-Wu model, a 2D Ising model with bond disorder perfectly correlated in one
dimension \cite{McCoyWu}. However, it was fully understood only when Fisher \cite{Fisher9295}
solved the 1D random transverse field Ising model by a version of the Ma-Dasgupta-Hu real space
renormalization group \cite{MDH79}. Since then, several infinite-randomness critical points
have been identified, mainly at quantum phase transitions since the disorder, being perfectly
correlated in (imaginary) time, has a stronger effect for quantum phase transitions than for
thermal ones. Examples include 1D random quantum spin chains as well as 1D and 2D random
quantum Ising models \cite{Bhatt82,Fisher94,Young96,Pich98,Motrunich00}.


\section{Griffiths singularities}
\label{sec:griffiths}

In the last section, we have discussed the global (average) behavior of the disorder at a
critical point. In this section, we focus on the effects of rare strong spatial disorder
fluctuations. Such fluctuations can lead to very interesting non-perturbative effects not only
directly at the phase transition but also in its vicinity. These effects, which are known as
Griffiths phenomena, can be understood as follows: Generically, the critical temperature $T_c$
of a disordered system is lower than its clean value, $T_c^0$. In the temperature interval
$T_c<T<T_c^0$, the bulk system is in the disordered phase. However, in an infinite size sample,
there is an exponentially small, but nonzero probability for finding an arbitrary large region
devoid of impurities. Such a rare region (or Griffiths island) can develop local order while
the bulk system is still disordered. The dynamics of a rare region is very slow because
flipping it requires a coherent change of the order parameter over a large volume. Griffiths
\cite{Griffiths69} showed that the presence of these locally ordered islands produces an
essential singularity \cite{Griffiths69,Bray89} in the free energy in the whole region
$T_c<T<T_c^0$, which is now known as the Griffiths region or the Griffiths phase
\cite{Randeria85}. In generic classical systems the Griffiths singularity is weak, and it does
not significantly contribute to the {\em thermodynamic} observables. To the best of our
knowledge, the thermodynamic Griffiths singularities in classical systems have not yet been
observed in experiments. In contrast, the long-time dynamics is dominated by the rare regions.
Inside the Griffiths phase, the spin autocorrelation function $C(t)$ decays as $\ln C(t)\sim
-(\ln t)^{d/(d-1)}$ for Ising systems \cite{Randeria85,Dhar,Dhar88,Bray88a,Bray88b} and as $\ln
C(t)\sim -t^{1/2}$ for Heisenberg systems \cite{Bray88a,Bray87}.

In quantum systems or classical systems with perfectly correlated disorder (like the
above-mentioned McCoy-Wu model), the Griffiths effects are generically stronger.
Effectively, the rare regions are now extended objects which are infinite in the
correlated space or imaginary time directions. This makes their dynamics even slower and so
increases their effects \cite{McCoy69}.
 In 1D and 2D random
transverse field Ising models, the singularity of the free energy in the Griffiths phase takes
a power-law form with nonuniversal continuously varying exponents. Several thermodynamic
observables including the average susceptibility actually diverge in a finite region of the
disordered phase \cite{Fisher9295,Young96,Pich98,Motrunich00} rather than only at the critical
point. Similar phenomena have also been found in quantum Ising spin glasses
\cite{ThillHuse95,gbh96,RiegerYoung96}. Because quantum Griffiths singularities are much
stronger than classical ones they should be accessible in experiments. Indeed, the peculiar
behavior of many heavy fermion systems in the vicinity of a magnetic quantum phase transition
has been interpreted as quantum Griffiths behavior \cite{CastroNetoJones} even though the
situation is probably more complex, at least for Ising symmetry \cite{MillisMorrSchmalian}, as
will be explained in the next section.


\section{Disorder induced smearing of a phase transition}
\label{sec:smearing}

\subsection{Smearing Mechanism}
In the preceeding sections, we have assumed that the phase transition remains sharp in the
presence of quenched disorder. We now show that in some systems the rare regions effects can
become so strong that the transition is completely destroyed by smearing.
Consider a single rare region that is locally in the ordered phase. In the generic case of
uncorrelated disorder it is of finite extension. Therefore it cannot undergo a true phase
transition independently of the bulk system. Instead, it very slowly fluctuates leading to the
Griffiths effects discussed in section \ref{sec:griffiths}. This argument suggests that the
phenomenology of the transition can completely change, if the system permits rare regions that
can undergo a true phase transition independently of the bulk, leading to a static order
parameter on the rare regions. It was recently found
\cite{rounding_prl,planar_ising,planar_mc,contact_us} that this does indeed happen, and that it
leads to a smeared global phase transition.

The behavior of such a smeared transition is markedly different from that of a conventional
(sharp) continuous phase transition. At a conventional transition, a non-zero order parameter
develops as a collective effect of the entire system which is signified by a diverging
correlation length of the order parameter fluctuations at the critical point and accompanying
singularities in the thermodynamic observables. In contrast, at a smeared transition the system
divides itself up into spatial regions which independently undergo the transition at different
values of the control parameter. Once true static order has developed on some of the rare
regions their order parameters can be aligned by an infinitesimally small interaction or
external field. Therefore, the global order develops very inhomogeneously over a
range of control parameter values. The correlation length remains finite, and the singularities
of the thermodynamic observables are rounded. To show the ubiquity of this smeared transition
scenario we now discuss three different examples.

\subsection{Magnetic quantum phase transitions of itinerant electrons}

The first class of systems displaying smeared transitions are itinerant quantum magnets with
Ising order parameter symmetry. For definiteness we consider the antiferromagnetic quantum
phase transition. The Landau-Ginzburg-Wilson free energy functional of the clean transition
reads \cite{Hertz76,BelitzKirkpatrick96}
\begin{equation}
S = \int dx\,dy\ m(x)\,\Gamma(x,y)\,m(y)
    + u\int dx\ m^4(x)~.
\label{eq:action}
\end{equation}
Here $m$ is the staggered magnetization, $x\equiv ({\bf x},\tau)$ comprises position ${\bf x}$
and imaginary time $\tau$, and $\int dx \equiv \int d{\bf x}\int_{0}^{1/T}d\tau$. $\Gamma(x,y)$
is the bare two-point vertex, whose Fourier transform is $\Gamma({\bf q},\omega_n) = (t + {\bf
q}^2 + \vert\omega_n\vert)$. Here $t=(g-g_c)/g_c$ is the distance of the coupling constant $g$
from the (clean) critical point. The dynamical part of $\Gamma$ is proportional to $|\omega_n|$
reflecting the overdamping of the dynamics due to the coupling of the order parameter to
fermionic particle-hole excitations (undamped dynamics leads to $\omega_n^2$). Quenched
disorder is introduced by making $t$ a random function of position, $t \to t + \delta t({\bf
x})$. We consider a Poisson (dilution) type of disorder, $\delta t({\bf x})=0$ everywhere
except on randomly distributed finite-size islands (impurities) of spatial density $p$ where
$\delta t({\bf x})=W>0$.

The rare regions in this system are large spatial regions devoid of these impurities. If $t<0$,
some of the rare regions are locally in the ordered phase. The crucial difference of the
itinerant system considered here and the transverse field Ising models of section
\ref{sec:griffiths} is the overdamped dynamics. The linear frequency dependence in $\Gamma$ is
equivalent to a long-range interaction in imaginary time of the form $(\tau-\tau')^{-2}$. Each rare
region is thus equivalent to an 1D Ising model with a $1/r^2$ interaction.  This model is known
to have a phase transition \cite{Ising1r2}. Thus, true static order can develop on those rare
regions which are locally in the ordered phase, leading to a smeared global phase transition as
described above \cite{rounding_prl}. Physically, this means the rare regions become static
because the damping prevents them from tunneling, as was independently shown in Ref.\
\cite{MillisMorrSchmalian}.

\subsection{Classical magnets with extended defects}

A second class of systems which can show smeared transitions are classical magnets with
extended impurities. As an example, consider a 3D classical Ising model with planar defects,
given by the Hamiltonian
\begin{eqnarray}
\label{eq:2} H=-\sum_{\genfrac{}{}{0pt}{}{i=1,\dots,L_\bot}{ j,k=1,\dots,L_C}
}J_iS_{i,j,k}S_{i+1,j,k} - \sum_{\genfrac{}{}{0pt}{}{i=1,\dots,L_\bot}
{j,k=1,\dots,L_C}}J(S_{i,j,k}S_{i,j+1,k}+S_{i,j,k}S_{i,j,k+1})~.
\end{eqnarray}
Here, $S_{ijk}=\pm 1$ are classical Ising spins on the sites of a cubic lattice. In the clean
system all interactions are identical and have the value $J$.  The defects are modelled via
'weak' bonds randomly distributed in one dimension, the uncorrelated dimension. The bonds in
the remaining two (correlated) dimensions remain equal to $J$.  $L_\bot$ and $L_C$ are the system
lengths in the uncorrelated and correlated directions, respectively. $J_i$ is the random
coupling constant in the uncorrelated direction, drawn from a binary probability distribution:
\begin{equation}
\label{eq:3} P[J_i] = (1-p)\, \delta(J_i-J) + p\, \delta(J_i-cJ)
\end{equation}
where $p$ and $c$ are constants between 0 and 1. Thus, the Ising magnet has planar impurities
of concentration $p$ and relative strength $c$ of the weak bonds.

Rare regions in this system consist of large blocks of planes devoid of weak bonds. They are
infinite in two dimensions (the correlated dimensions) but finite in the uncorrelated
direction. Since a 2D Ising model has an ordered phase, each rare region can independently
undergo the magnetic phase transition. This leads to a smeared global phase transition in the
3D Ising model with planar defects \cite{planar_ising,planar_mc}. Physically, the reason for
the smearing is different from the itinerant magnets discussed above. Here, an isolated rare
region can develop a static order parameter because its dimensionality is above the lower
critical dimension $d_c^-=1$ for the classical Ising model. This also shows that a continuous
order parameter symmetry suppresses the tendency towards smearing. In classical magnets with XY
or Heisenberg symmetry and short-range interactions, the defects must be at least
three-dimensional for smearing to occur since the lower critical dimension for these systems is
$d_c^-=2$.

\subsection{Nonequilibrium transitions}
A third class of smeared phase transitions are non-equilibrium phase transitions with extended
spatial defects. Consider, e.g., the contact process \cite{contact}, a prototypical system in
the directed percolation universality class \cite{dp}. It is defined on a $d$-dimensional
hypercubic lattice. Each site can be vacant or active, i.e, occupied by a particle. During the
time evolution, particles are created at vacant sites at a rate $\lambda n/ (2d)$ where $n$ is
the number of active nearest neighbor sites and the `birth rate' $\lambda$ is the control
parameter. Particles are annihilated at unit rate. For small $\lambda$, annihilation dominates,
and the absorbing state without any particles is the only steady state. For large $\lambda$
there is a steady state with finite particle density (active phase). Both phases are separated
by a nonequilibrium phase transition in the directed percolation universality class at $\lambda=\lambda_c^0$. We
introduce quenched spatial disorder by making the birth rate $\lambda$ a random function of the
lattice site. As in the preceeding section, extended impurities can be described by disorder
perfectly correlated in $d_C$ dimensions, but uncorrelated in the remaining $d_R=d-d_C$
dimensions. $\lambda$ is thus a function of ${\mathbf r}_R$ which is the projection of the
position vector $\mathbf r$ on the uncorrelated directions. For definiteness, we assume that
the $\lambda({\mathbf r}_R)$ have a binary probability distribution
\begin{equation}
P[\lambda({\mathbf r}_R)] = (1-p)\, \delta[\lambda({\mathbf r}_R)-\lambda] + p\,
\delta[\lambda({\mathbf r}_R) - c\lambda]
\end{equation}
where $p$ and $c$ are constants between 0 and 1. In other words, there are extended impurities
of density $p$ where the birth rate $\lambda$ is reduced by a factor $c$.

Rare regions in this disordered contact process are infinite in the correlated dimensions $d_C
\ge 1$ but finite in the $d_R$ random directions. The directed percolation universality class
permits a phase transition for all dimensions $d\ge 1$. Therefore, each rare region can undergo
a phase transition independently of the bulk system. Following the general arguments above, we
thus conclude that the global phase transition of a system in the directed percolation
universality class is smeared by extended spatial defects \cite{contact_us}.

\subsection{Extremal statistics theory}

In this section we derive the leading behavior in the tail of a smeared phase transition, i.e.,
in the parameter region where a few isolated rare regions have already ordered, but their
distance is so large that they are effectively independent. For definiteness we consider the
first example, a magnetic quantum phase transition of itinerant electrons. Adaption to the
other discussed systems is straight forward. The theory is similar to that of Lifshitz
\cite{Lifshitz} and others for the band tails in a disordered semiconductor.

The probability $w$ for finding a region of linear size $L_{RR}$ devoid of any impurities is
given by $w \sim \exp( -p L_{RR}^d)$ (up to pre-exponential factors). Such a rare region
develops static order at a distance $t_c(L_{RR}) < 0$ from the {\em clean} critical point.
Finite size scaling yields $|t_c(L_{RR})| \sim L_{RR}^{-\phi}$ where $\phi$ is the finite-size
scaling shift exponent of the clean system \cite{shiftexponent}. Thus, the probability for
finding a rare region which becomes critical at $t_c$ is given by
\begin{equation}
w(t_c) \sim \exp  (-B ~|t_c|^{-d/\phi}) \qquad \textrm{for } t\to 0-~. \label{eq:wtc}
\end{equation}
The total order parameter $m$ is obtained by integrating over all rare regions which are
ordered at $t$, i.e., all rare regions having $t_c>t$. This leads to an exponential tail of $m$
as a function of the distance $t$ from the {\em clean} critical point:
\begin{equation}
\log (m) \sim -B |t|^{-d/\phi}  \qquad \textrm{for } t\to 0-~. \label{eq:m-dilute}
\end{equation}
Note that the functional dependence on $t$ of the order parameter on a given island is of
power-law type and thus only influences the pre-exponential factors.

At finite temperatures, the static order on the rare regions is destroyed, and a finite
interaction of the order of the temperature is necessary to align them. This means a sharp
phase transition is recovered. To estimate the transition temperature we note that the
interaction between two rare regions depends exponentially on their spatial distance $r$ which
itself, according to (\ref{eq:wtc}), depends exponentially on $t$. This leads to a
double-exponential dependence of the critical temperature on $t$
\begin{eqnarray}
 \log( -a \log T_c) &\sim& |t|^{-d/\phi} ~.
 \end{eqnarray}

We now turn to finite-size effects at zero temperature. The total order parameter is the sum of
contributions of many independent islands. Therefore, finite size-effects in a macroscopic
sample are governed by the central limit theorem. However, for $t\to 0-$, very large and thus
very rare regions are responsible for the order parameter. The number $N$ of rare regions which
start to order at $t_c$ in a sample of size $L$ behaves like $N \sim L^d ~w(t_c)$. When $N$
becomes of order one, strong sample-to-sample fluctuations arise. Using eq.\ (\ref{eq:wtc}) for
$w(t_c)$, we find that sample-to-sample fluctuations become important at
\begin{eqnarray}
|t_L| &\sim& (\log L)^{-\phi/d} . \label{eq:tL-dilute}
\end{eqnarray}

The spatial magnetization distribution in the tail of the smeared transition is very
inhomogeneous. On the ordered rare regions, the local order parameter $m({\bf r})$ is
comparable to that of the clean system. Away from these islands, it decays exponentially with
distance. The probability distribution $P[\log m({\bf r})]$ is therefore very broad, ranging
from $\log m({\bf r}) = O(1)$ on the largest islands to $\log m({\bf r}) \to -\infty$ on sites
very far away from an ordered island. The {\em typical} local order parameter $m_{typ}$ can be
estimated from the typical distance of any point to the nearest ordered island. From
(\ref{eq:wtc}) we obtain
\begin{equation}
r_{typ} \sim \exp  (B ~|t|^{-d/\phi}/d) ~. \label{eq:rtyp}
\end{equation}
At this distance from an ordered island, the local order parameter has decayed to
\begin{equation}
m_{typ} \sim e^{-r_{typ}/\xi_0} \sim \exp \left[ -C \exp(B ~|t|^{-d/\phi}/d)\right]~
\label{eq:mtyp}
\end{equation}
where $\xi_0$ is the bulk correlation length (which is finite and changes slowly throughout the
tail region of the smeared transition)  and $C$ is constant. A comparison with
(\ref{eq:m-dilute}) gives the relation between $m_{typ}$ and the thermodynamic (average) order
parameter $m$,
\begin{equation}
|\log m_{typ}| \sim m^{-1/d} \label{eq:mtypav}~.
\end{equation}
Thus, $m_{typ}$ decays exponentially with $m$ indicating an extremely broad local order
parameter distribution. In order to determine the functional form of the local order parameter
distribution, first consider a situation with just a single ordered island at the origin of the
coordinate system. For large distances ${\bf r}$ the magnetization falls off exponentially like
$m({\bf r}) = m_0~ e^{-r/\xi_0}$. The probability distribution of $x=\log[m({\bf r})]=\log m_0
-r/\xi_0$ can be calculated from
\begin{equation}
P(|x|) = \left |\frac {dN}{dx} \right| = \frac{dN}{dr} \left | \frac {dr}{dx}\right | =\xi_0
\frac{dN}{dr}
       \sim \xi_0 r^{d-1}
\label{eq:probdis}
\end{equation}
where $dN$ is the number of sites at a distance from the origin between $r$ and $r+dr$ or,
equivalently, having a logarithm of the local magnetization between $x$ and $x+dx$. For large
distances we have $|x| \sim r$. Therefore, the probability distribution of $\log m$ generated
by a single ordered island takes the form
\begin{equation}
 P[\log(m)] \sim |\log(m)|^{d-1}   \qquad (\textrm{for } m \ll 1)~.
 \label{eq:plogm}
\end{equation}
In the tail region of the smeared transition, the system consists of a few ordered islands
whose distance is large compared to $\xi_0$. The probability distribution of $\log[m({\bf r})]$
thus takes the form (\ref{eq:plogm}) with a lower cutoff corresponding to the typical
island-island distance and an upper cutoff corresponding to a distance $\xi_0$ from an ordered
island.


\section{Numerical results}
\label{sec:numerics}

In this section, we illustrate the disorder-induced smearing of a phase transition by showing
numerical data for a classical Ising model with two spatial and one time-like dimensions. The
disorder is totally correlated in the time-like direction. The interactions are short-ranged in
space but infinite-ranged in the time-like direction. This simplification retains the crucial
property of static order on the rare regions but permits system sizes large enough to study
exponentially rare events. The Hamiltonian reads
\begin{equation}
H= - \frac 1 L_\tau\sum_{\langle{\bf x,y}\rangle,\tau,\tau'}   S_{{\bf x},\tau} S_{{\bf
y},\tau'}
   - \frac 1 L_\tau\sum_{{\bf x  },\tau,\tau'} J_{\bf x} S_{{\bf x},\tau} S_{{\bf x},\tau'}~.
\label{eq:toy}
\end{equation}
Here ${\bf x}$ and $\tau$ are the space and time-like coordinates, respectively. $L_\tau$ is
the system size in time direction and $\langle{\bf x,y}\rangle$ denotes pairs of nearest
neighbors. $J_{\bf x}$ is a quenched binary random variable with the distribution $P(J) =
(1-p)~ \delta(J-1) + p~ \delta(J)$. In the following, we fix the impurity concentration at
$p=0.2$, unless otherwise noted. In this classical model, $L_\tau$ takes the role of the
inverse temperature in the corresponding quantum system and the classical temperature takes the
role of the coupling constant $g$. Because the interaction is infinite-ranged in time, the
time-like dimension can be treated in mean-field theory. For $L_\tau\to\infty$, this leads to a
set of coupled mean-field equations (one for each {\bf x})
\begin{equation}
 m_{\bf x} = \tanh \beta~ [ J_{\bf x} m_{\bf x} + {\sum_{{\bf y}({\bf x})}}' m_{\bf y} + h]~,
\label{eq:mf}
\end{equation}
where $h=10^{-8}$ is a small symmetry-breaking magnetic field. Eqs. (\ref{eq:mf}) are solved
numerically in a self-consistency cycle.

\begin{figure}
\includegraphics[width=7.5cm]{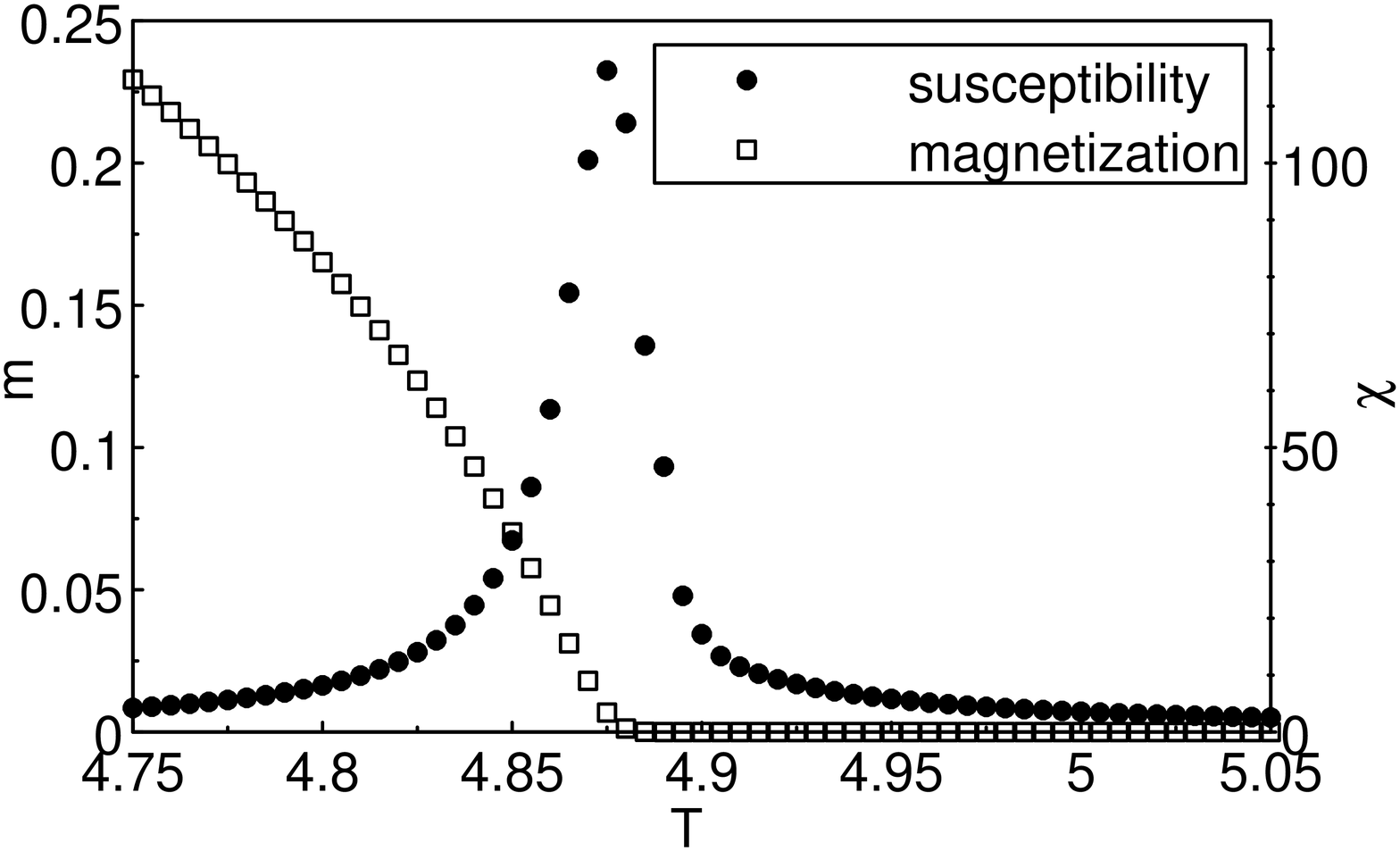}
\includegraphics[width=7.8cm]{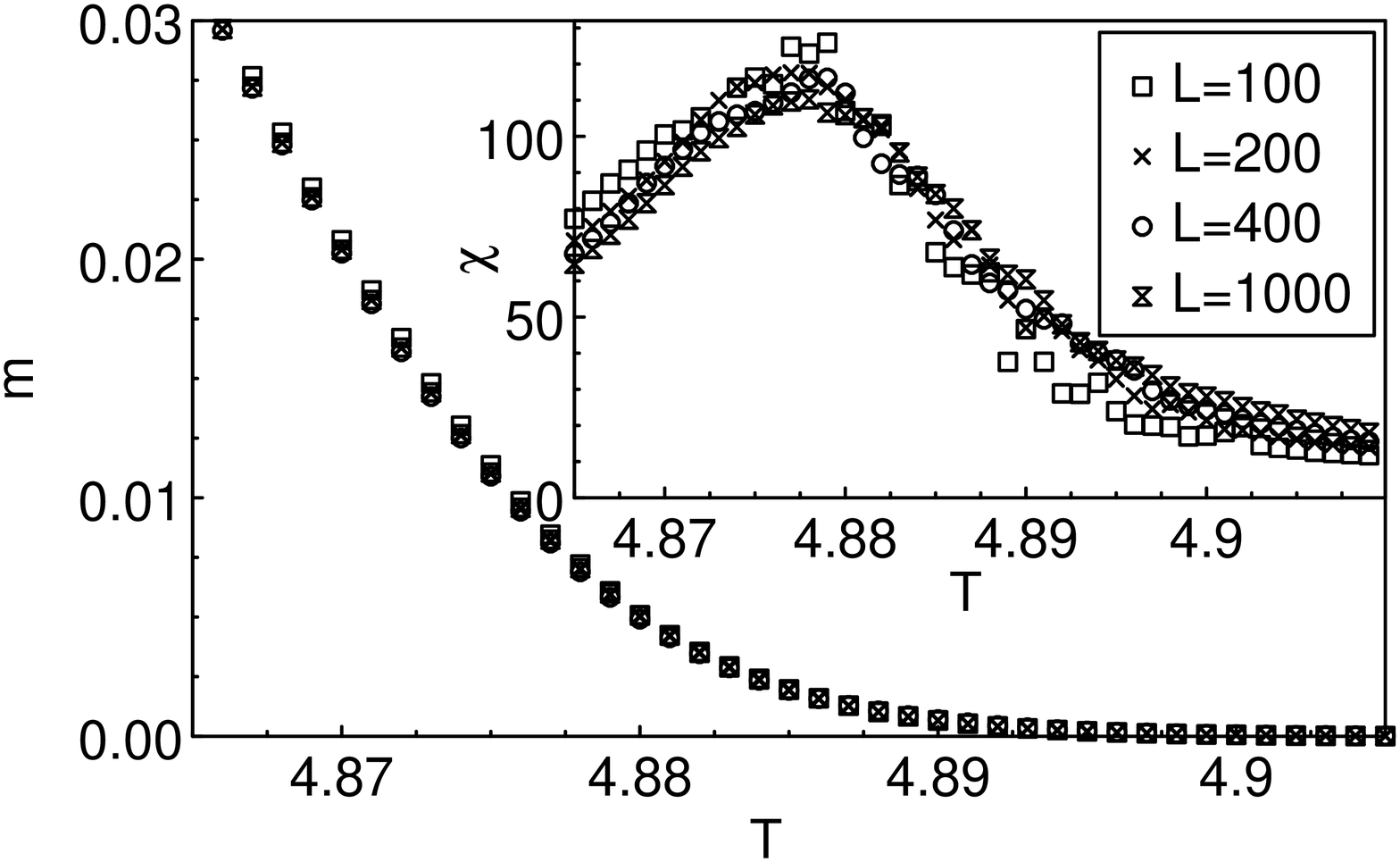} \caption{Magnetization and susceptibility in the vicinity
of the smeared transition. Left: Overview of the behavior for  $L=100$ and $p=0.2$. There is a
seeming transition at $T\approx 4.88$. Right: The tail region of the smeared transition. The
agreement of the results of different systems sizes shows that the observed rounding is not a
finite-size effect (from reference \protect\cite{rounding_prl}).} \label{Fig:overview}
\end{figure}

The left panel of Fig.\ \ref{Fig:overview} shows the total magnetization and the susceptibility (corresponding
to the inverse energy gap of the quantum system) as functions of temperature for linear size
$L=100$. The data are averages over 200 disorder realizations \cite{average}.
At a first glance these data suggest a sharp phase transition at $T\approx 4.88$. However, a
closer investigation (right panel of Fig.\ \ref{Fig:overview}) shows that the singularities
are rounded. If this rounding was a
conventional finite-size effect the magnetization curve should become sharper with increasing
$L$ and the susceptibility peak should diverge. This is not the case here. Instead, the
transition is intrinsically smeared for $L\to\infty$.

\begin{SCfigure}
\includegraphics[width=8cm]{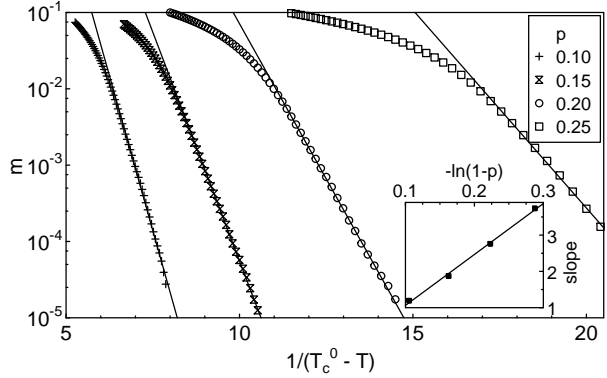} \caption{Log($m$) as a function of the distance
from the clean critical point for four different impurity concentrations.
         The solid lines are fits of the average magnetization to eq.\ (\ref{eq:m-dilute}) with $\phi=2$.
         The inset shows  the dependence of the slope of the curves on $-\ln(1-p)$.}
\label{Fig:logarithm}
\end{SCfigure}

For comparison with the analytical results, Fig.\ \ref{Fig:logarithm} shows the logarithm of
the average magnetization as a function of $1/(T_{c}^0-T)$ for four different impurity
concentrations. Here, $T_{c}^0=5$ is the critical temperature of the clean system ($p=0)$. All
data sets follow eq.\ (\ref{eq:m-dilute}) over several orders of magnitude in $m$ with the
expected shift exponent of $\phi=2$. The slope of each curve gives the corresponding prefactor
$B$ in eq.\ (\ref{eq:m-dilute}). In our lattice model, $B$ should be proportional to
$-\ln(1-p)$. The inset of Fig.\ \ref{Fig:logarithm} shows that this relation is fulfilled in
good approximation.

\begin{SCfigure}
\includegraphics[width=10.5cm]{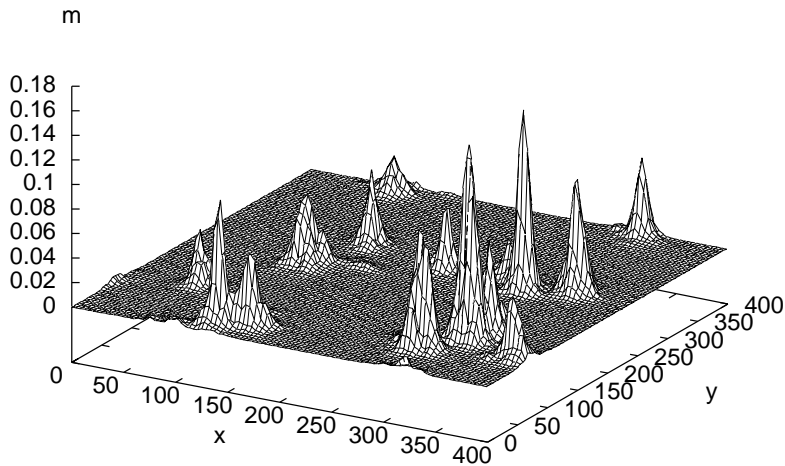}
\caption{Local magnetization of a single sample of linear size  $L=400$ in the random
directions for $T=4.8875$.} \label{Fig:locmag}
\end{SCfigure}

\begin{SCfigure}
\includegraphics[width=8.cm]{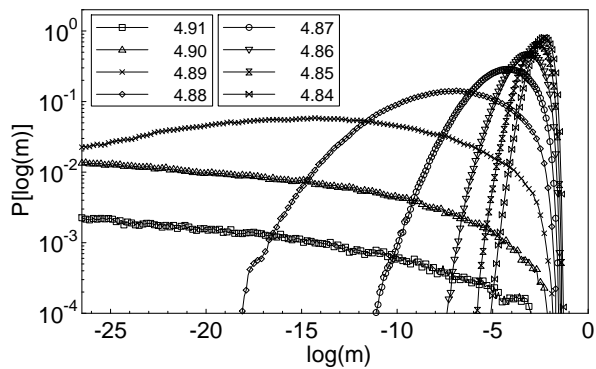}
\caption{Probability distribution of the local magnetization of a single sample of linear size
$L=2000$ in the random directions for several temperatures.} \label{Fig:distrib}
\end{SCfigure}

In Fig.\ \ref{Fig:locmag} we demonstrate the extreme inhomogeneity of the system in the tail of
the smeared transition. It shows the local magnetization of a single sample of linear size
$L=400$ for $T=4.8875$ as a function of the spatial coordinates. Clearly, a sizable
magnetization only exists on a few isolated islands; in-between it is vanishingly small. Figure
\ref{Fig:distrib} shows the probability distribution of the local magnetization for several
temperatures, calculated from a single sample of linear size $L=2000$. In the tail of the
smeared transition, i.e., for temperatures close to $T_c^0=5$, the distribution becomes
extremely broad, even on a logarithmic scale. A more detailed investigation
\cite{planar_ising,Sknepnek} shows that this distribution indeed follows eqs.\
(\ref{eq:mtypav}) and (\ref{eq:probdis}).


\section{Summary and conclusions}
\label{sec:conclusions}

To summarize, we have shown that in several disordered classical and quantum systems, isolated
rare spatial regions can undergo a true phase transition independently of the bulk system. As a
consequence, different parts of the system undergo the phase transition at different control
parameter values, i.e, the global phase transition is smeared by the disorder. For classical
phase transitions, this scenario requires extended defects whose dimensionality is above the
lower critical dimension for the universality class in question. Specifically, for classical
Ising and Heisenberg magnets the (equilibrium) phase transition is smeared if the
dimensionalities of the defects are at least two and three, respectively. Non-equilibrium phase
transitions in the directed-percolation universality class are smeared if the defects are at
least linear. Quantum phase transitions display a greater tendency towards smearing because
quenched disorder is always perfectly correlated in imaginary time direction (which has to be
taken into account at quantum phase transitions). At magnetic quantum phase transitions in
itinerant electron systems, the rare region effects are further enhanced by Landau damping.
This leads to smearing even for point-like impurities.
In this final section we discuss questions of universality, the relation to Griffiths
phenomena,
and the Harris criterion. We also discuss favorable conditions for observing the smearing in
experiments or simulations.

The origins of the disorder-induced smearing of the phase transition described in this paper
and of Griffiths phenomena are very similar, both are caused by rare large spatial regions
which are locally in the ordered phase. The difference lies in the dynamics of the rare
regions. In the case of Griffiths effects, the order parameter on a rare region fluctuates
slowly leading to the singularities \cite{Griffiths69} discussed in Sec.\ \ref{sec:griffiths}.
In contrast, smearing occurs if the rare regions actually develop true static order.

A second remark deals with the relation of the disorder-induced smearing and the Harris
criterion. We emphasize that the phase transition can be smeared by planar defects even if the
corresponding clean critical point fulfills the Harris criterion and appears to be stable. The
reason is that the Harris criterion assumes a {\em homogeneous} transition and studies the
behavior of the coarse-grained (root-mean-square) disorder at large length scales. However, the
formation of static order on an isolated finite-size rare region is a non-perturbative
finite-length scale effect in the tail of the disorder distribution. This type of effects is
not covered by the Harris criterion.

Let us also note that the functional dependence of the magnetization and other observables on
the temperature is {\it not} universal, it depends on details of the disorder distribution.
Therefore, only the presence or absence of smearing is universal in the sense of critical
phenomena (i.e., depending on dimensionality and symmetry only) while the thermodynamic
relations are non-universal.

We now turn to favorable conditions for the {\it observation} of the smearing in experiments or
simulations. A significant number of large rare regions will only exist if the defect
concentration is small. On the other hand, the impurities have to be sufficiently strong so
that the bulk system is still far away from criticality when the first rare regions start to
order. Thus, the most favorable type of disorder for the observation of the smearing is a small
density of strong impurities.

\begin{acknowledgement}
This paper is dedicated to Prof.\ Michael Schreiber on the occasion of his 50th birthday.
Michael Schreiber has made several important contribution to the physics of disordered systems
ranging from Anderson localization and interaction effects to hopping transport and Coulomb gap
physics. This work was supported in part by the University of Missouri Research Board.
\end{acknowledgement}


\begin{thebibliography}{99}
\frenchspacing
\bibitem{Grinstein} G. Grinstein {\it Fundamental Problems in Statistical
 Mechanics VI}, ed E.G.D. Cohen (Elsevier, New York, 1985) p.147
\bibitem{Harris74} A.B. Harris, J. Phys. C {\bf 7}, 1671 (1974)
\bibitem{Fisher9295} D.S. Fisher, Phys. Rev. Lett. {\bf 69} 534, (1992); Phys. Rev. B {\bf 51}, 6411 (1995)
\bibitem{McCoyWu} B.M. McCoy and T.T. Wu, Phys. Rev. {\bf 176}, 631 (1968); Phys. Rev. {\bf
        188}, 982 (1969)
\bibitem{rounding_prl} T. Vojta, Phys. Rev. Lett. {\bf 90}, 107202 (2003)
\bibitem{AharonyHarris96} A. Aharony and A.B. Harris,  Phys. Rev. Lett. {\bf 77}, 3700 (1996)
\bibitem{WisemanDomany98} S. Wiseman and E. Domany, Phys. Rev. Lett. {\bf 81}, 22 (1998)
\bibitem{Janke93} C. Holm and W. Janke, Phys. Rev. B {\bf 48}, 936 (1993)
\bibitem{Ferrenberg91} A.M. Ferrenberg and D.P. Landau, Phys. Rev. B {\bf 44}, 5081 (1991)
\bibitem{Ballesteros98} H.G. Ballesteros et al. , Phys. Rev. B {\bf 58}, 2740 (1998)
\bibitem{MDH79} S.K. Ma, C. Dasgupta and C.-K. Hu, Phys. Rev. Lett. {\bf 43}, 1434 (1979)
\bibitem{Bhatt82} R.N. Bhatt and P.A. Lee, Phys. Rev. Lett. {\bf 48}, 344 (1982)
\bibitem{Fisher94} D.S. Fisher, Phys. Rev. B {\bf 50}, 3799 (1994)
\bibitem{Young96}A.P. Young and H. Rieger, Phys. Rev. B {\bf 53}, 8486 (1996)
\bibitem{Pich98} C. Pich, A. P. Young, H. Rieger, and N. Kawashima, Phys. Rev. Lett. {\bf 81},
        5916 (1998)
\bibitem{Motrunich00} O. Motrunich, S.-C. Mau, D.A. Huse, and D.S. Fisher, Phys. Rev. B {\bf 61},
        1160(2000)
\bibitem{Griffiths69} R.B. Griffiths, Phys. Rev. Lett. {\bf 23}, 17 (1969)
\bibitem{Bray89} A.J. Bray  and D. Huifang, Phys. Rev. B {\bf 40}, 6980 (1989)
\bibitem{Randeria85} M. Randeria, J. Sethna, and R.G. Palmer, Phys. Rev. Lett. {\bf 54}, 1321
        (1985)
\bibitem{Dhar} D. Dhar, {\it Stochastic Processes: Formalism and Applications}, ed D.S. Argawal
   and S. Dattagupta (Berlin, Springer, 1983)
\bibitem{Dhar88} D. Dhar, M. Randeria, and J.P. Sethna, Europhys. Lett. {\bf 5}, 485 (1988)
\bibitem{Bray88a} A.J. Bray, Phys. Rev. Lett. {\bf 60}, 720 (1988)
\bibitem{Bray88b} A.J. Bray and G.J. Rodgers, Phys. Rev. B {\bf 38}, 9252 (1988)
\bibitem{Bray87} A.J. Bray, Phys. Rev. Lett. {\bf 59}, 586 (1987)
\bibitem{McCoy69}B.M. McCoy, Phys. Rev. Lett. {\bf 23}, 383 (1969)
\bibitem{ThillHuse95}M. Thill and D. Huse, Physica A {\bf 214}, 321 (1995)
\bibitem{gbh96}M. Guo, R. Bhatt and D. Huse, Phys. Rev. B {\bf 54}, 3336 (1996)
\bibitem{RiegerYoung96}H. Rieger and A.P. Young, Phys. Rev. B {\bf 54}, 3328 (1996)
\bibitem{CastroNetoJones}A.H. Castro Neto, G. Castilla, and B.A. Jones, Phys. Rev. Lett. {\bf 81}, 3531 (1998);
        A.H. Castro Neto and B.A. Jones, Phys. Rev. B {\bf 62}, 14975 (2000)
\bibitem{MillisMorrSchmalian} A.J. Millis, D.K. Morr, and J. Schmalian, Phys. Rev. Lett. {\bf 87}, 167202
       (2001); Phys. Rev. B {\bf 66}, 174433 (2002)
\bibitem{planar_ising} T. Vojta, J. Phys. A {\bf 36}, 10921 (2003)
\bibitem{planar_mc} R. Sknepnek and T. Vojta, Phys. Rev. B in print, cond-mat/0311394
\bibitem{contact_us} T. Vojta, cond-mat/0402606
\bibitem{Hertz76} J. Hertz, Phys. Rev. B {\bf 14}, 1165 (1976)
\bibitem{BelitzKirkpatrick96} T.R. Kirkpatrick and D. Belitz, Phys. Rev. Lett. {\bf 76}, 2571 (1996);
        {\bf 78}, 1197 (1997)
\bibitem{Ising1r2}D.J. Thouless, Phys. Rev. {\bf 187}, 732 (1969);
        J. Cardy, J. Phys. A {\bf 14}, 1407 (1981)
\bibitem{contact} T.E. Harris, Ann. Prob. {\bf 2}, 969 (1974)
\bibitem{dp} P. Grassberger and A. de la Torre, Ann. Phys. (NY) {\bf 122}, 373
    (1979).
\bibitem{Lifshitz} I.M. Lifshitz, Usp. Fiz. Nauk {\bf 83}, 617 (1964)
                  [Sov. Phys.--Usp. {\bf 7}, 549 (1964)]
\bibitem{shiftexponent}The upper critical dimension of the clean itinerant antiferromagnetic quantum
     phase transition is $d_c^+=2$. Thus, hyperscaling is not valid in 3D and $\phi \ne 1/\nu$. For our
     purpose the exact value of $\phi$ is not important.
\bibitem{average}Thermodynamic quantities involve averaging over the whole system. Thus ensemble averages
     rather than typical values give the correct infinite system
     approximation.
\bibitem{Sknepnek} R. Sknepnek and T. Vojta, unpublished

\end{thebibliography}
\end{document}